\documentstyle[fleqn]{tp}

\input epsf

\def\ltsima{$\; \buildrel < \over \sim \;$}
\def\lsim{\lower.5ex\hbox{\ltsima}}
\def\gtsima{$\; \buildrel > \over \sim \;$}
\def\gsim{\lower.5ex\hbox{\gtsima}}

\begin{document}

\twocolumn[
\title{Simulating Weak Lensing by Clusters and Large-Scale Structure}
\author{Bhuvnesh Jain$^1$, Uros Seljak$^2$ and Simon White$^2$\\
{\it $^1$Dept. of Physics and Astronomy, Johns Hopkins University, 
Baltimore MD, USA}\\
{\it $^2$Max-Planck-Institut f\"ur Astrophysik, D--85740 Garching}}
\vspace*{16pt}   

ABSTRACT.\
Selected results on estimating cosmological parameters from simulated weak 
lensing data with noise are presented. Numerical simulations of ray tracing 
through N-body simulations have been used to generate shear and convergence
maps due to lensing by large-scale structure. Noise due to the intrinsic
ellipticities of a finite number of galaxies is added. In this contribution
we present our main results on estimation of the power spectrum and density 
parameter $\Omega$ from weak lensing data on several degree sized fields. We 
also show that there are striking morphological differences in the
weak lensing maps of clusters of galaxies formed in models with 
different values of $\Omega$. 

\endabstract]

\markboth{Jain et al.}{Numerical simulations of weak lensing}

\small

\section{Introduction}
Weak lensing by large-scale structure (LSS) shears the images of distant
galaxies. 
The first calculations of weak lensing by LSS (Blandford et al. 1991; 
Miralda-Escude 1991; Kaiser 1992), based on the pioneering work of
Gunn (1967), showed that lensing would induce coherent ellipticities
of order 1$\%$ over regions of order one degree on the sky. Recently several
authors have extended this work to probe semi-analytically the 
possibility of measuring  the mass power spectrum and cosmological parameters
from the second and third moments of the induced ellipticity or 
convergence (Bernardeau et al. 1997; Kaiser 1998; Stebbins 1996; Jain and Seljak 1997; Schneider et al. 1997).

Analytical calculations suggest that nonlinear evolution of
the density perturbations that provide the lensing effect can significantly
alter the predicted signal. It is expected to enhance 
the power spectrum on small scales
and makes the probability distribution function (pdf) of
the ellipticity and convergence non-Gaussian. We have carried out
numerical simulations of ray tracing through N-body simulation data to
compute the fully nonlinear moments and pdf. Details of the method
and results are presented in a separate (Jain, Seljak \& White 1999); 
here we summarize the method and present some highlights of the results. 
Other recent numerical work includes Wambsganss, Cen \& Ostriker (1998),
Premadi, Martel \& Matzner (1998), van Waerbeke, Bernardeau \& Mellier (1998),
Bartelmann et al (1998) and Couchman, Barber \& Thomas (1998)

The dark matter distribution obtained from N-body simulations of
different models of structure formation is projected on to 2-dimensional
planes lying between the observer and source galaxies. Typically we 
use galaxies at $z\sim 1$ with
$\sim 20-30$ planes. We propagate $\sim 10^6$ light rays through these planes
by computing the deflections due to the matter at every plane. 
Fast Fourier Transforms are used to compute gradients of the potential
that provide the shear tensor at each plane. The outcome
of the simulation is a map of the shear and convergence on square
patches of side length $1-5^{\circ}$. Several realizations for each model
are needed to compute reliable statistics on scales ranging from $1'$
to 1$^{\circ}$. 

\begin{figure*}[t]
\vspace*{13cm}
\caption{The dimensionless power spectrum of $\kappa$. 
For the cosmological model indicated by $\Omega_m$ and
$\Gamma$ in the panel, the power spectrum from ray tracing shown by the
solid curves is
compared with the linear (long-dashed) and nonlinear analytical 
(short-dashed) predictions. 
The angular wavenumber $l$ is given in inverse radians -- the smallest
$l$ plotted corresponds to modes with wavelength of order 
$L\simeq 3^{\circ}$, where $L$ is the side-length of the field. 
}
\includegraphics{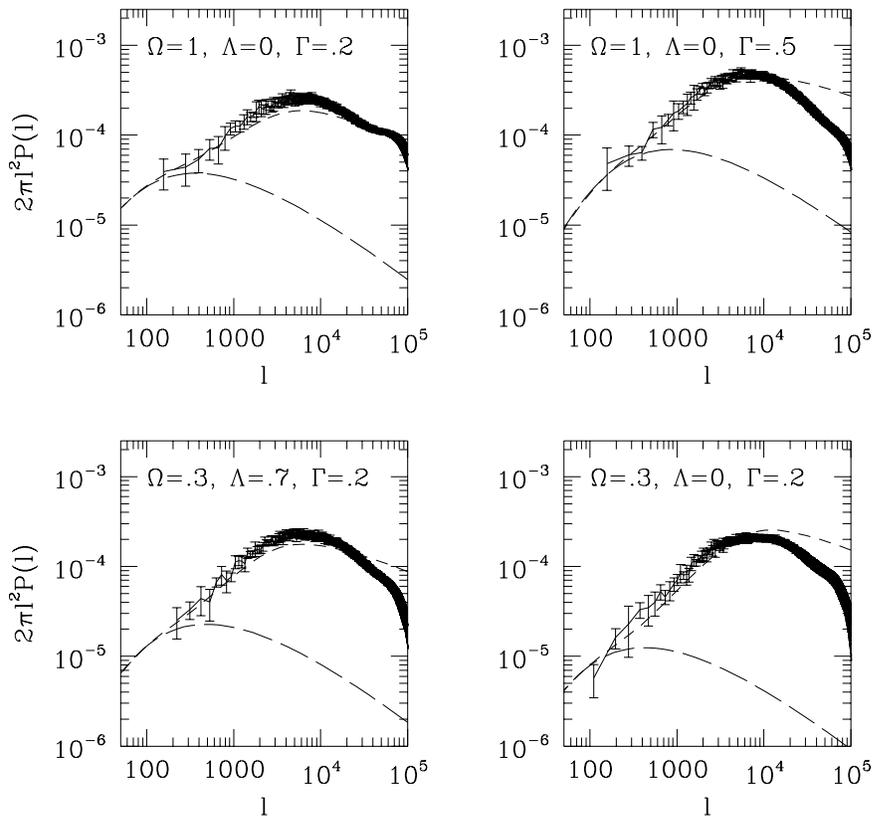}
\label{figpower}
\end{figure*}

In the weak lensing regime, the magnification and induced ellipticity
are given by linear combinations of the Jacobian matrix of the mapping
from the source to the image plane. The Jacobian matrix is defined by
\begin{equation}
\Phi_{ij} \equiv  {\partial \delta \theta_i \over \partial \theta_j}
\end{equation}
where $\delta \theta_i$ is the $i-$th component of the perturbation due to
lensing of the angular position on the 
source plane, and $\theta_j$ is the $j-$th component of the position on 
the image plane. The convergence is defined as 
$\kappa=-(\Phi_{11}+\Phi_{22})/2$, while
the two components of the shear are $\gamma_1=-(\Phi_{11}-\Phi_{22})/2$ 
and $\gamma_2=-\Phi_{12}$. The convergence $\kappa$ can be reconstructed
from the measured shear $\gamma_1$, $\gamma_2$, up to a constant which 
depends on the mean density in the survey area. If the survey is 
sufficiently large and there is little power on scales larger than the
survey, this error can be neglected.

\section{Results of Simulations}

The N-body simulations used for the ray tracing 
are taken from four different cosmological
simulations, the parameters of which are summarized in 
Table~\ref{tab:1}.  The N-body simulations use an
adaptive $P^3M$  method with $256^3$ particles, and were carried out
using codes kindly made available by the Virgo consortium (e.g. 
Jenkins et al. 1997). 
Coupled with ray tracing on a $2048^2$ grid these simulations 
provide us with a small scale 
resolution down to $\lsim 0.5'$, well into the nonlinear regime for weak
lensing. 

\begin{table*}
\caption[]{Summary of the parameters used for the N-body
  simulations. $h$ is the Hubble constant in units of $100\,{\rm
  km\,s^{-1}\,Mpc^{-1}}$, $\Gamma$ is the shape parameter of the power
  spectrum, and the other parameters have their conventional meaning.}
  \centering
  \begin{tabular}{l|*{7}{c}}
    \hline
    Model & $\Omega_m$ & $\Omega_\Lambda$ & $h$ & $\sigma_8$ &
    $\Gamma$ \\
    \hline
    SCDM         & 1.0 & 0.0 & 0.5 & 0.60 & 0.50 \\
    $\tau$CDM    & 1.0 & 0.0 & 0.5 & 0.60 & 0.21 \\
    $\Lambda$CDM & 0.3 & 0.7 & 0.7 & 0.90 & 0.21 \\
    OCDM         & 0.3 & 0.0 & 0.7 & 0.85 & 0.21 \\
    \hline
  \end{tabular}
\label{tab:1}
\end{table*}

\begin{figure*}[ht!]
\vspace*{13.5cm}
\caption{The reconstructed convergence field from noisy
ellipticity data on a single 3$^\circ$ field. 
The upper left panel shows the field without noise or sparse sampling. 
The upper right panel shows the reconstructed field using
$2\times 10^5$ galaxies per square degree at a mean $z=1$. 
The lower left panel used $1/4$th as many galaxies.  
The lower right panel shows a
map of  the ``curl'' field generated from the
same ellipticity data as in the lower left panel. Fluctuations in
this field are solely due to noise, therefore
comparison of the two panels shows the significance of the 
reconstructed features in the lower left panel. 
}
\includegraphics{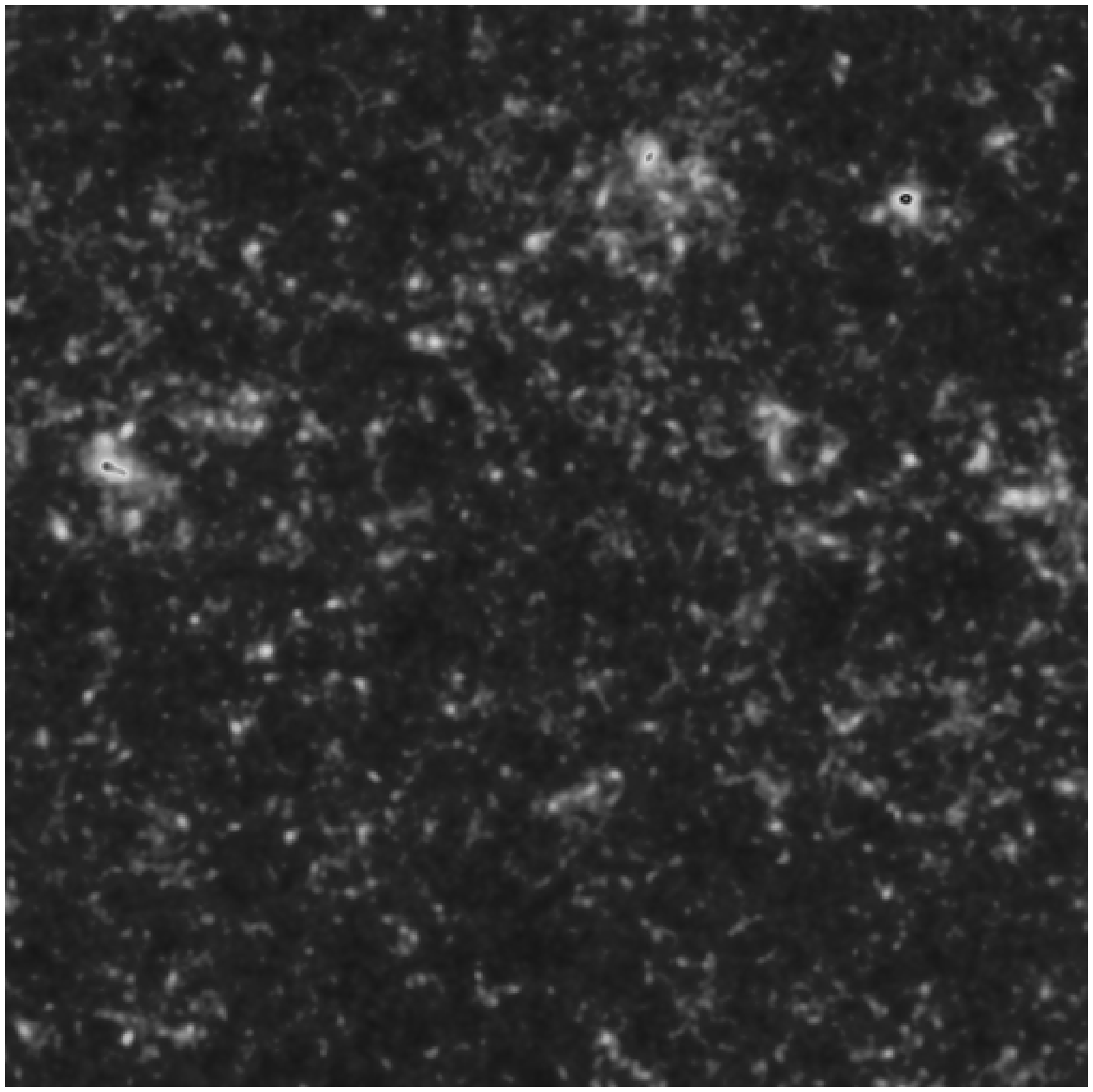}
\includegraphics{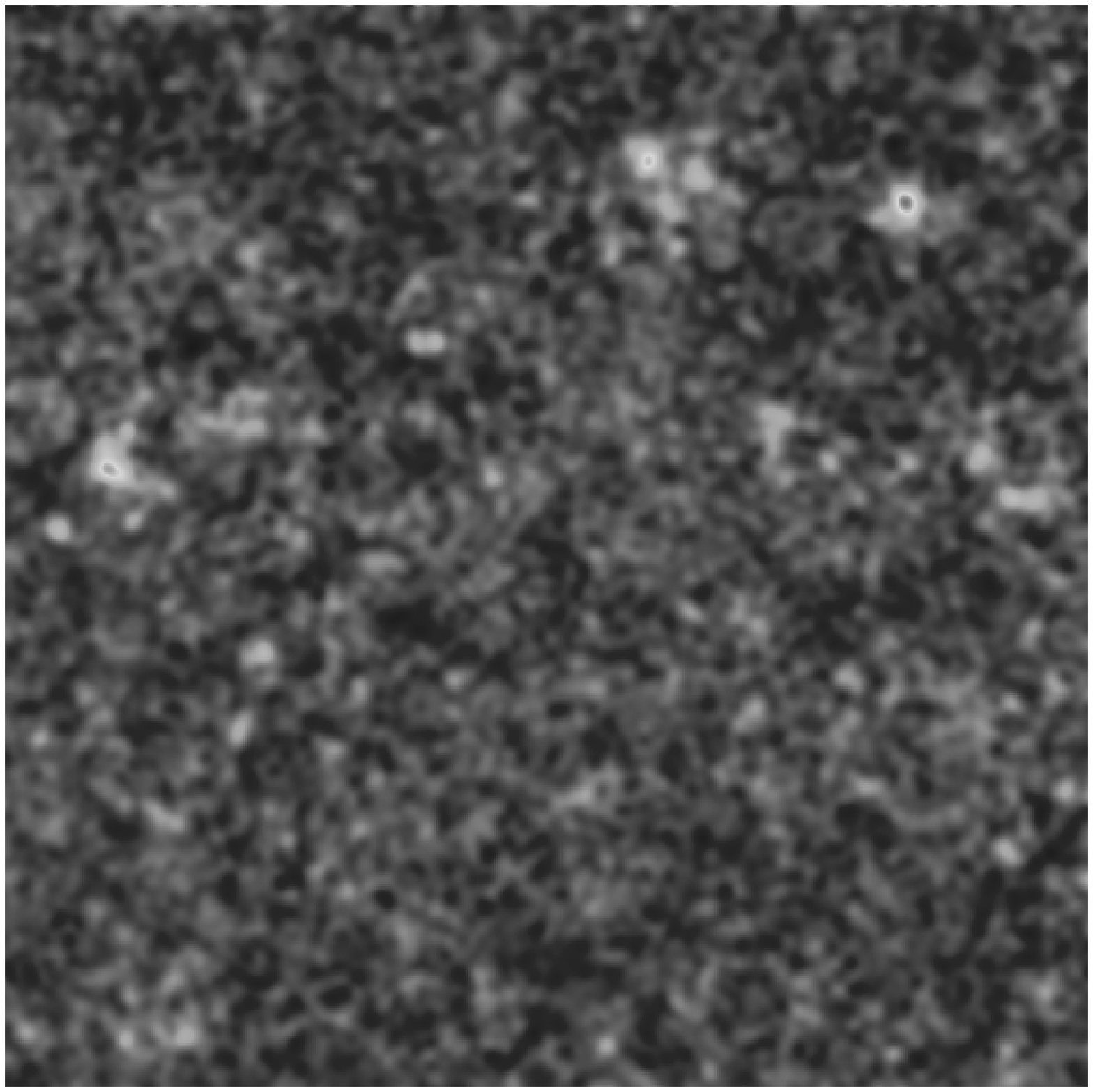}
\includegraphics{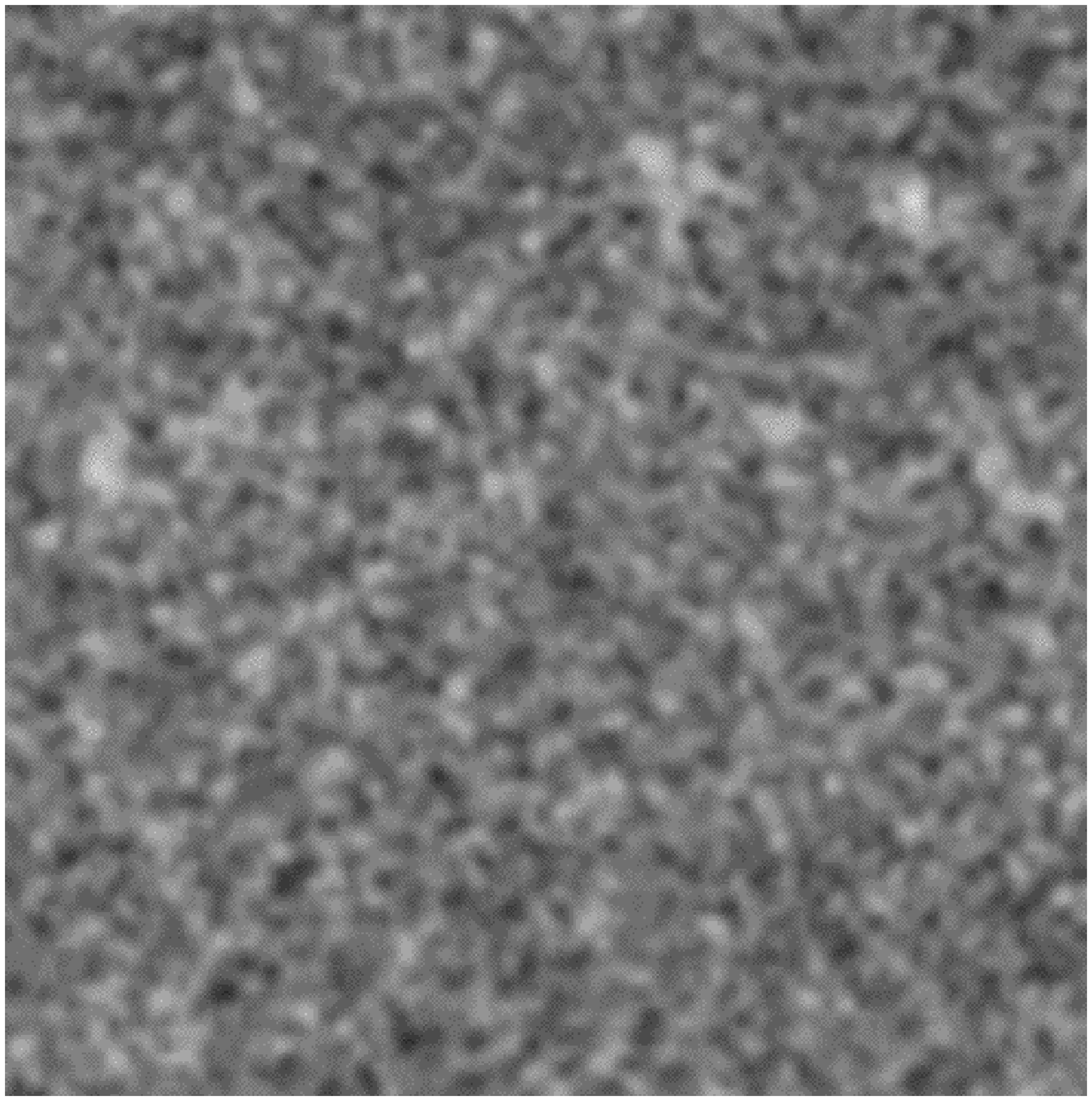}
\includegraphics{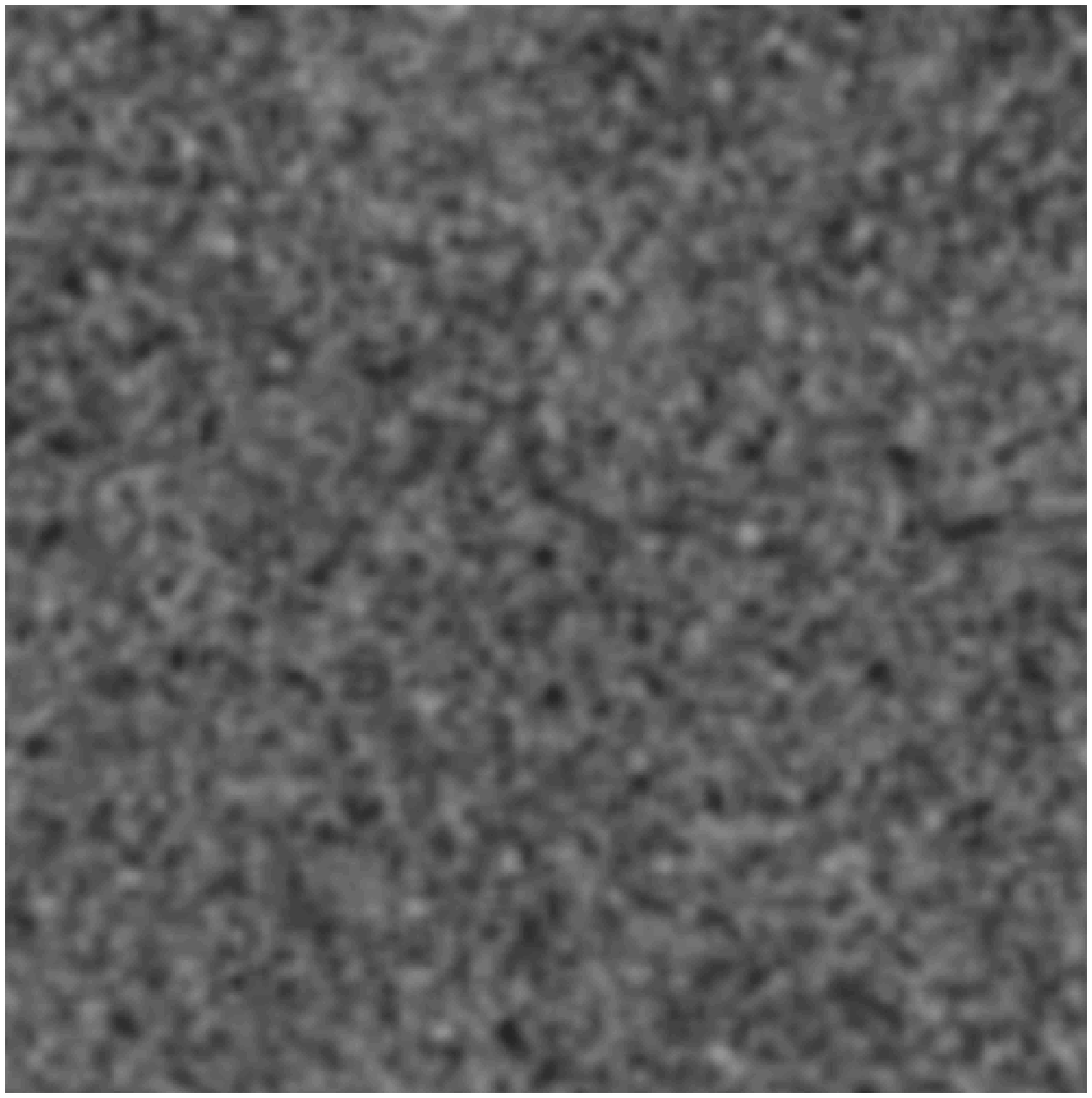}
\label{figkappa}
\end{figure*}

The power spectrum of $\kappa$ measured from the simulations
is compared with the analytical predictions of Jain \& Seljak (1997)
in figure \ref{figpower}. We will focus in this contribution on 
the extraction of parameters from simulated noisy data. Figure \ref{figkappa}
shows the reconstructed convergence field from noisy
ellipticity data on a single 3$^\circ$ field. The galaxies 
were randomly distributed and assigned intrinsic
ellipticites with each component drawn randomly from a Gaussian with
rms=0.4. The reconstructed $\kappa$ is compared with the field without
noise, as well as with a pure noise field, to show the significance
of reconstructed features. The field is smoothed on the scale at which
the noise and signal should be comparable. The figure shows that with
observational parameters feasible on large telescopes, a field a few
degrees on a side can allow one to reconstruct the large-scale features
dominated by groups and clusters of galaxies. Statistically one can
do much better. 

\begin{figure*}[htb!]
\vspace*{9cm}
\caption{Convergence power spectrum estimated from simulated noisy data. 
The shear field on a single field 3$^\circ$ on a side is sampled
by randomly distributed galaxies with intrinsic ellipticites
assigned as in figure \ref{figkappa}. The power spectrum of the 
reconstructed $\kappa$ from the ellipticity data is shown with error
bars obtained from 10 independent realizations of the noisy data. 
}
\includegraphics{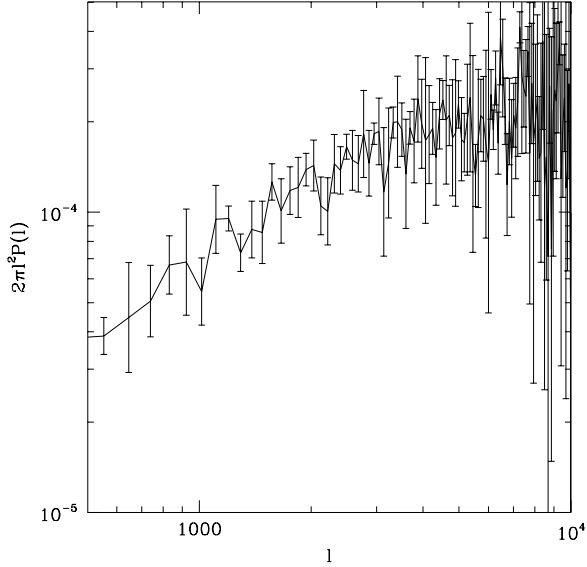}
\label{figpowernoise}
\end{figure*}

Figure \ref{figpowernoise} shows the power spectrum of $\kappa$ 
measured from simulated noisy
data. On scales larger than 10' the signal dominates the noise. 
The scales on which the power can be measured 
are dominated by the density power spectrum on 
scales of about $1-10\ h^{-1}$Mpc at $z\sim 0.3$. Thus weak lensing
surveys a few degrees on a side will be sensitive to the dark matter
power spectrum on these scales.

\begin{figure*}[htb!]
\vspace*{9cm}
\caption{The density parameter $\Omega_m$ estimated from the pdf of
simulated noisy data. The skewness parameter
$S_3$ is estimated by minimizing the $\chi^2$ with respect to
an Edgeworth expansion of the pdf. The four curves with
decreasing peak heights are for the open, cosmological constant, 
Einstein-de Sitter models, all with $\Gamma=0.2$ CDM power spectra, 
and an Einstein-de Sitter model with $\Gamma=0.5$ CDM power 
spectrum. 
}
\includegraphics{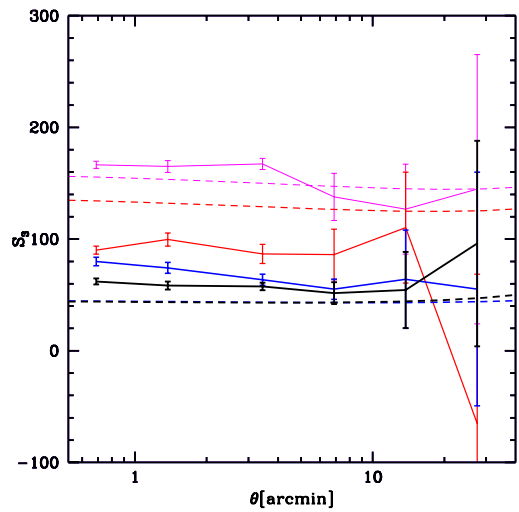}
\label{fignoise}
\end{figure*}

Figure \ref{fignoise} shows one attempt at estimating the density
parameter $\Omega_m$ from simulated data. The simplest way to estimate
$\Omega_m$ is to measure the skewness $S_3$ of the convergence 
(Bernardea et al. 1997; Jain \& Seljak 1997; Schneider et al. 1998). 
We found that $S_3$ was sensitive to the tails of the probability
distribution function (pdf) and therefore required large sample sizes. 
Instead, we fit the pdf to an Edgeworth expansion and estimated $S_3$ 
as a parameter. The result is shown in figure \ref{fignoise} and shows
that cosmological models with different values of $\Omega_m$ can be
distinguished at a high level of significance (the error bars are 1-$\sigma$). 
The dashed curves show the predictions of perturbation theory. 

\begin{figure*}[ht!]
\vspace*{13cm}
\caption{Clusters in open and Einstein-de Sitter cosmologies. 
The convergence in fields 10' on a side
centered on a rich cluster is shown for the EdS model in the
upper panels and for the open model in the lower panels. 
The values of the convergence range from over 10\% in the
center to below 1\% in the outermost regions. Figure \ref{figprofile}
gives the mean profiles of the convergence. 
}
\includegraphics{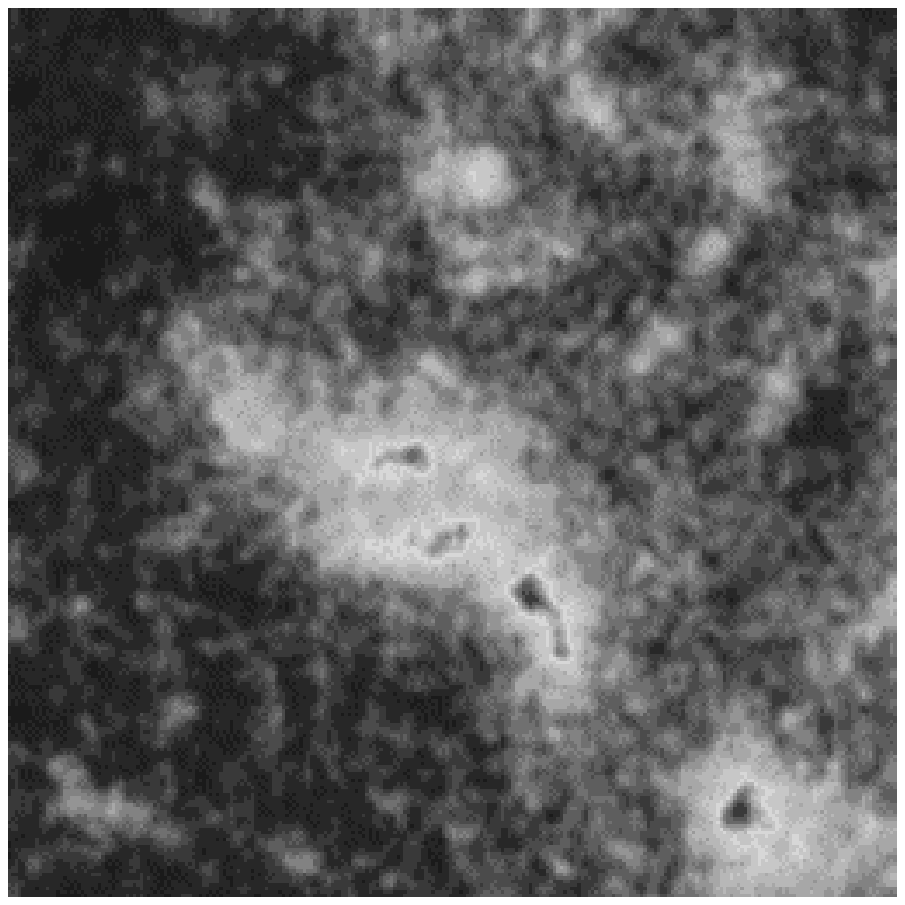}
\includegraphics{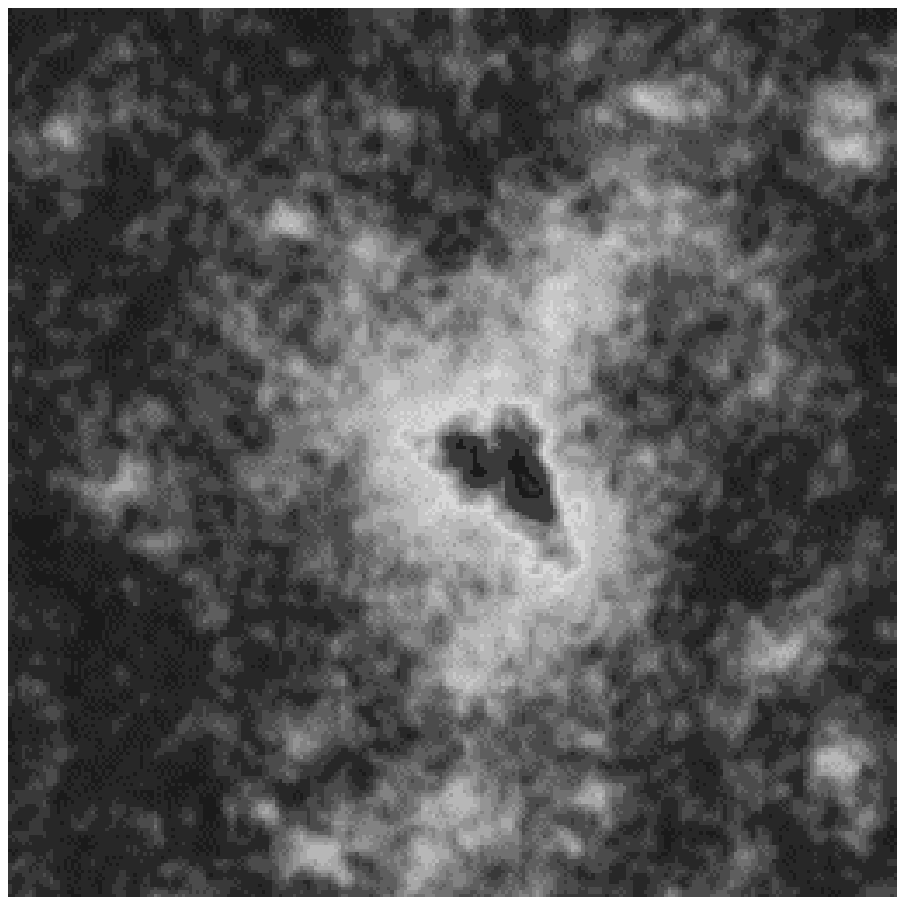}
\includegraphics{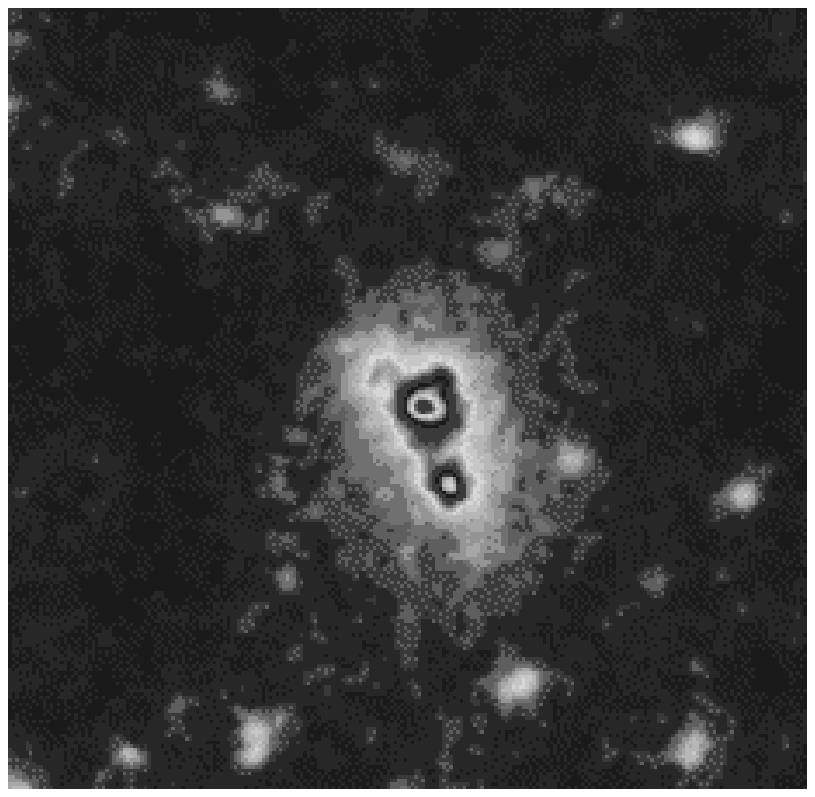}
\includegraphics{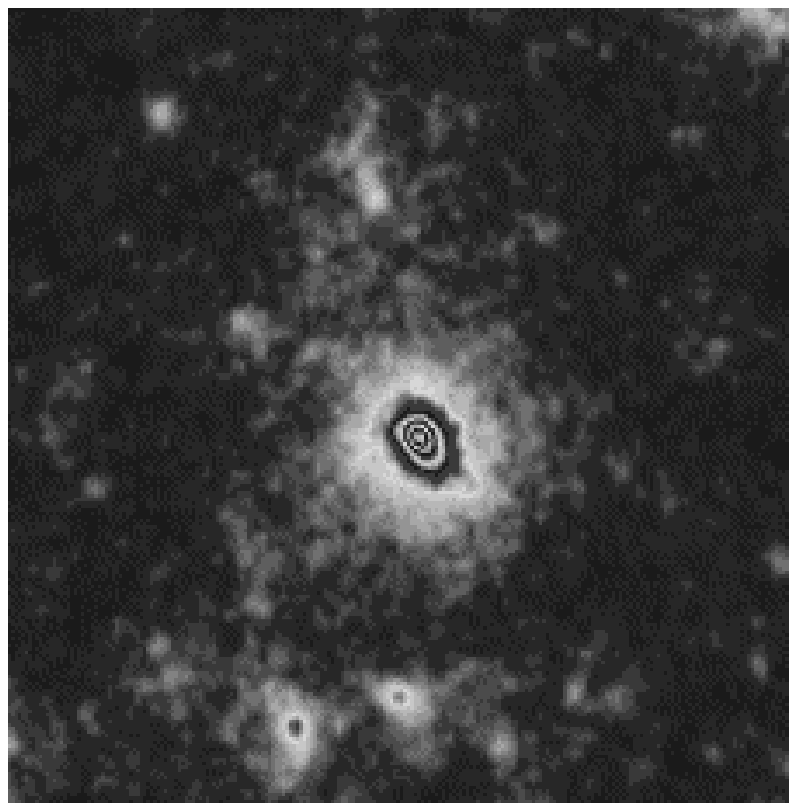}
\label{figclus}
\end{figure*}

\begin{figure*}[htb!]
\vspace*{7cm}
\caption{The profile of the convergence in clusters in open
and EdS cosmologies. The dashed curve shows the average convergence 
profile measured from 10 clusters in the EdS model while the solid 
curve shows the corresponding profile for the open model. 
}
\includegraphics{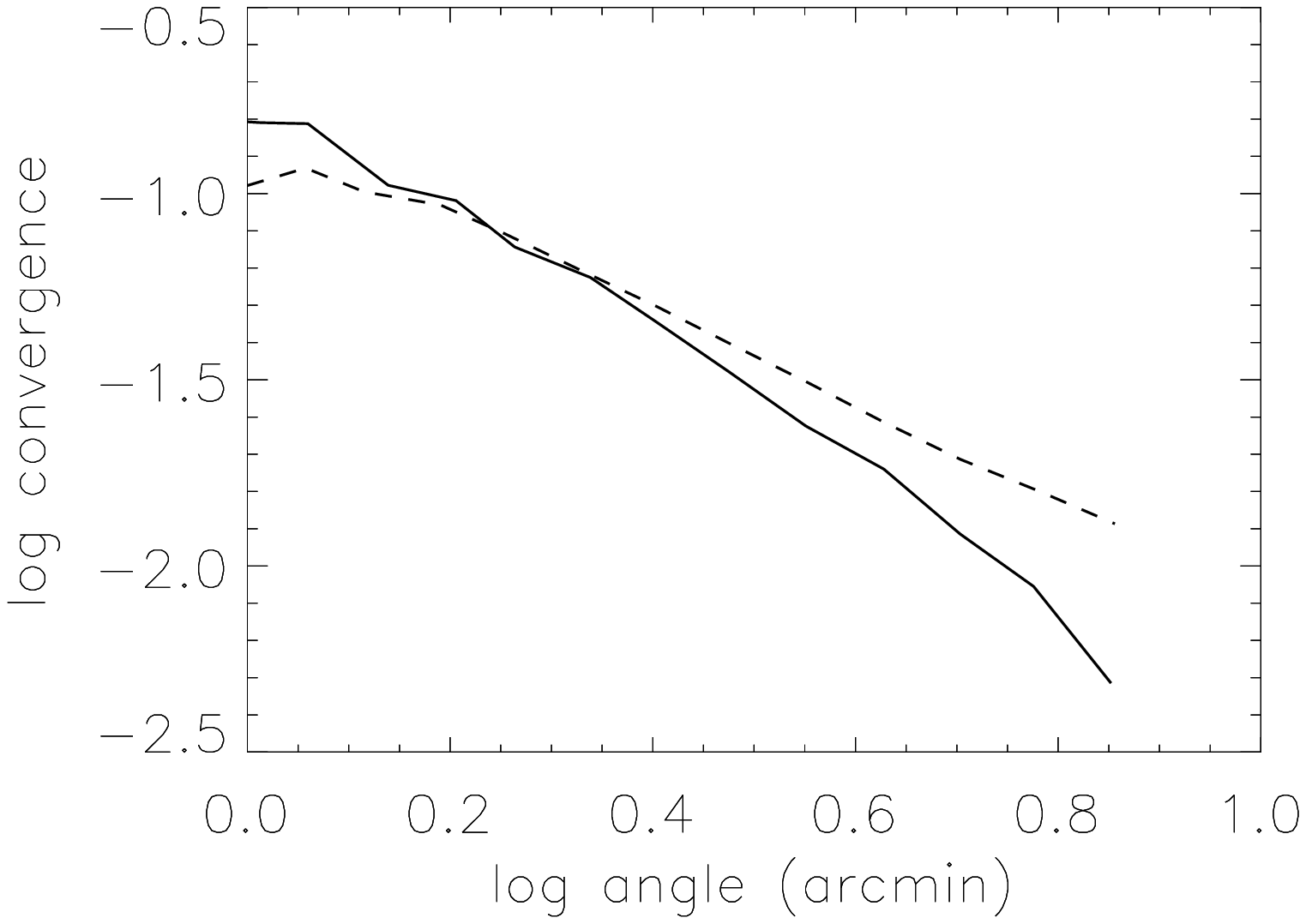}
\label{figprofile}
\end{figure*}

Finally, figure \ref{figclus} shows the convergence field centered
on typical rich clusters in the Einstein-de Sitter model, $\tau$CDM, 
and the open CDM model. The fields are 10' arcminutes on a side and
are chosen because current observational data is already capable of 
detecting signal in the outer parts of clusters. Comparison of the 
EdS and open models shows that the convergence fields are morphologically
different: clusters in the open model appear more compact, regular and
isolated. The clusters in the EdS model are not fully relaxed and appear
linked to the surrounding large-scale structure through filaments or
more irregular structures. These differences can be quantified by 
using topological measures and by measuring moments such as the quadrupole
or higher moments. In figure \ref{figprofile} we simply show the mean 
profile of the convergence around clusters, which also shows differences
between the two models. The source galaxies have been taken
to be at $z=1$ and the 10 richest clusters in a field 3 degrees on a 
side have been used to obtain the mean profiles. 

\section{Conclusion}

We have shown results on estimating the power spectrum and
$\Omega_{\rm m}$ from simulated, noisy weak lensing data. 
With several fields a degree on a side, the dark matter power
spectrum on scales of $1-10\ h^{-1}$Mpc can be probed and
$\Omega_{\rm m}$ estimated to within about 0.1-0.2. We have
also shown preliminary results on the morphological differences
between clusters in different cosmologies. Further work is needed
to quantify the differences and explore the effects of noise and 
of varying the redshift distribution of source galaxies. 

\section*{Acknowledgments}

We are grateful to Matthias Bartelmann and Peter Schneider
for helpful discussions. It is a pleasure to thank Anthony Banday
for his patient support in the writing of this contribution. 

\end{document}